\pdfoutput=1

\documentclass[pdflatex,sn-mathphys-num]{sn-jnl}
\usepackage{graphicx}%
\usepackage{multirow}%
\usepackage{amsmath,amssymb,amsfonts}%
\usepackage{amsthm}%
\usepackage{mathrsfs}%
\usepackage[title]{appendix}%
\usepackage{xcolor}%
\usepackage{textcomp}%
\usepackage{manyfoot}%
\usepackage{booktabs}%
\usepackage{algorithm}%
\usepackage{algorithmicx}%
\usepackage{algpseudocode}%
\usepackage{listings}%
\usepackage{caption}%
\usepackage{siunitx}%
\usepackage{natbib}%

\theoremstyle{thmstyleone}%
\theoremstyle{thmstyletwo}%

\theoremstyle{thmstylethree}%
\begin{document}

\title[Article Title]{Physics-guided foundation model for universal speckle removal in ultrathin multimode fiber imaging}


\author[1,2]{\fnm{Xianrui} \sur{Zeng}}

\author[3]{\fnm{Yirui} \sur{Zang}}

\author[1]{\fnm{Pengfei} \sur{Liu}}

\author[1,2]{\fnm{Fei} \sur{Yu}}

\author[3]{\fnm{Yang} \sur{Yang}}

\author[4,5,6]{\fnm{Tom\'a\v{s}} \sur{\v{C}i\v{z}m\'ar}}

\author*[1]{\fnm{Yang} \sur{Du}}\email{yangdu@ucas.ac.cn}

\affil[1]{\orgdiv{Hangzhou Institute for Advanced Study}, \orgname{University of Chinese Academy of Sciences}, \orgaddress{\city{Hangzhou}, \postcode{310024}, \country{China}}}

\affil[2]{\orgdiv{Shanghai Institute of Optics and Fine Mechanics}, \orgname{Chinese Academy of Sciences}, \orgaddress{\city{Shanghai}, \postcode{201800}, \country{China}}}

\affil[3]{\orgdiv{Department of Thoracic Surgery}, \orgdiv{Shanghai Pulmonary Hospital}, \orgdiv{School of Medicine}, \orgname{Tongji University}, \orgaddress{\city{Shanghai}, \postcode{200433}, \country{China}}}

\affil[4]{\orgdiv{Leibniz Institute of Photonic Technology}, \orgaddress{\street{Albert-Einstein-Stra{\ss}e 9}, \city{Jena}, \postcode{07745}, \country{Germany}}}

\affil[5]{\orgdiv{Institute of Scientific Instruments of CAS}, \orgaddress{\street{Kr\'alovopolsk\'a 147}, \city{Brno}, \postcode{61264}, \country{Czechia}}}

\affil[6]{\orgdiv{Institute of Applied Optics}, \orgname{Friedrich Schiller University Jena}, \orgaddress{\street{Fr\"obelstieg 1}, \city{Jena}, \postcode{07743}, \country{Germany}}}

\abstract{Ultrathin multimode fibers (MMFs) promise endoscopes with hair-scale diameters for accessing sub-millimeter anatomy, but in MMF far-field imaging the required small collection aperture drives speckle-dominated measurements that rapidly degrade image fidelity. Here we present Speckle Clean Network (SCNet), a physics-guided foundation model for universal speckle removal that makes photon-limited, single-fiber collection compatible with high-fidelity reconstruction across diverse scattering conditions without target-specific retraining. SCNet combines a Mixture of Experts (MoE) architecture with material-aware routing, wavelet-based frequency decomposition to separate structure from speckle across sub-bands, and a curriculum-style optimization that progressively enforces spectral consistency before spatial fidelity. Using an ultrathin dual-fiber holographic probe, we deliver wavefront-shaped illumination through one MMF and collect backscattered photons through a parallel MMF. We validate SCNet on 3D plastic objects over varying working distances, resolve 5.66 lp/mm on a paper USAF target, and restore fine structures on leaves and metal surfaces. On rabbit heart and kidney tissues, SCNet improves recovery of low-contrast anatomical texture under the same ultrathin collection constraint. We further compress SCNet through multi-teacher distillation to reduce computation while preserving reconstruction quality, enabling inference at 60 FPS. This work effectively decouples image quality from probe size, establishing a speckle-free ultrathin endoscopy for stand-off imaging in confined spaces.}




\maketitle
\section{Introduction}\label{sec1}

Endoscopic imaging provides a critical window into the body, enabling direct optical access to internal anatomy for visualizing tissue architecture and cellular dynamics in situ\cite{waterhouse2019roadmap}. 
While standard endoscopes are indispensable for large cavities, they are too bulky to navigate the body’s most delicate and inaccessible regions, such as peripheral microvessels, distal airways, and the intricate cochlear structures\cite{choi2022flexible,kennedy2022targeted,nyberg2019delivery}. Accessing these sub-millimeter cavities requires miniaturizing the imaging probe to hair-thin dimensions, a constraint that fundamentally conflicts with the optical requirements for high-fidelity imaging.

MMFs have emerged as a compelling solution to this dimensional constraint, offering a scalable platform for ultrathin endoscopy by guiding tens of thousands of modes within a footprint comparable to a single human hair\cite{vcivzmar2011shaping,vcivzmar2012exploiting,papadopoulos2012focusing,ploschner2015seeing}. 
By employing wavefront shaping to unscramble modal mixing, MMFs can function as high-resolution imaging tools, enabling minimally invasive access for deep brain fluorescence imaging and neuronal connectivity mapping\cite{turtaev2018high,ohayon2018minimally,vasquez2018subcellular,stibuurek2023110}. Recent advances have further enhanced these capabilities through dynamic wavefront control\cite{wang2025real}, memory effect exploiting\cite{li2021memory}, light-field encoding\cite{wen2023single}, and cascaded aberration correction\cite{wen2025cascaded} for improving imaging stability. 
Beyond contact or near-field operation, the field has extended to the observation of distant objects, achieved by generating diffraction-limited foci in the far field\cite{leite2021observing,stellinga2021time}. 
The far-field imaging expands the application space from surface-adjacent microendoscopy to macroscopic scene observation and depth-resolved interrogation at working distances set by anatomy rather than probe geometry. However, capturing sufficient backscattered signal from these distant targets typically necessitates a separate collection path. 
While integrating a large-core fiber can improve collection efficiency and reduce speckle contrast, this hardware modification inevitably increases the probe dimension, negating the primary advantage of fiber-based endoscopy. 
Conversely, collecting through a small-core fiber preserves a hair-thin footprint but yields speckle-dominated images with high contrast fluctuations that obscure fine structure and reduce reconstruction fidelity.

Parallel to these hardware constraints, deep learning has emerged as a transformative paradigm for overcoming physical limitations in computational imaging. In the broader field of image restoration, advanced architectures have evolved from residual convolutional networks\cite{zhang2017beyond,zamir2020learning,chen2021hinet}to powerful Transformers\cite{zamir2022restormer,zhang2023xformer,brateanu418akdt}and state-space models\cite{guo2024mambair,ma2024u}that capture long-range dependencies with unprecedented fidelity. 
This capability has been further refined by architectures that unify global and local modeling\cite{ning2025glnet}, employ iterative dynamic filtering networks\cite{kim2025idf}, or utilize adaptive integration of frequency mining and modulation\cite{cui2025adair} to enhance restoration accuracy. 
Recent innovations have expanded this toolkit even further with diffusion models\cite{cheng2024diffusion}, cross-spectral guidance\cite{wang2025complementary,sheng2023structure}, and self-supervised or zero-shot frameworks\cite{ma2025pixel2pixel,liu2023dynamic,laine2023high} that eliminate the need for paired training data. 
Within the specific domain of MMF imaging, data-driven approaches have been successfully adapted to model the complex light transmission, enabling robust image reconstruction through disordered modes\cite{rahmani2018multimode,matthes2021learning,fan2019deep} and maintaining performance under dynamic fiber perturbations\cite{resisi2021image,zhu2023anti}. 
Additionally, integrating physical operators into neural networks has enabled optical phase retrieval\cite{tu2025deep}and single shot wide-field imaging in reflection geometries\cite{liu2023single}. 
However, applying these models to the stochastic, high-contrast speckle noise inherent in single fiber far-field imaging presents a distinct challenge. 
Existing MMF-specific networks typically focus on transmission configurations or require extensive, material-specific calibration that does not generalize well to the diverse optical properties of distant targets. 
Conversely, general purpose denoisers often struggle to disentangle signal from the unique physics of modal interference speckle without explicit guidance, limiting their utility for real-time, high-fidelity endoscopic reconstruction.

Here we present SCNet, a physics-guided foundation model that enables universal speckle removal for ultrathin multimode fiber far-field imaging without material-specific retraining. 
In conjunction with a compact dual-fiber holographic probe (one fiber for wavefront-shaped illumination and a second co-aligned fiber for backscattered photon collection), SCNet integrates a MoE module with material-aware routing, a wavelet-based frequency-domain operator that helps disentangle speckle from structure, and a curriculum-style optimization that progressively enforces spectral consistency before spatial fidelity. 
We validate this framework across a broad range of scattering regimes and working distances, spanning plastic, paper, metal, plant leaves and biological tissues, and show that it recovers fine structure under photon-limited single-fiber collection, including 5.66 line pairs/mm resolution on paper and sub-millimeter anatomical features in rabbit organs. 
Finally, we compress SCNet via multi-teacher distillation to reduce computational cost while maintaining reconstruction quality, enabling inference 60 FPS for practical endoscopic deployment.

\section{Results}\label{sec2}

\subsection{Framework of SCNet}\label{subsec1}
\begin{figure}[htbp!]
\centering
\includegraphics[width=1\textwidth]{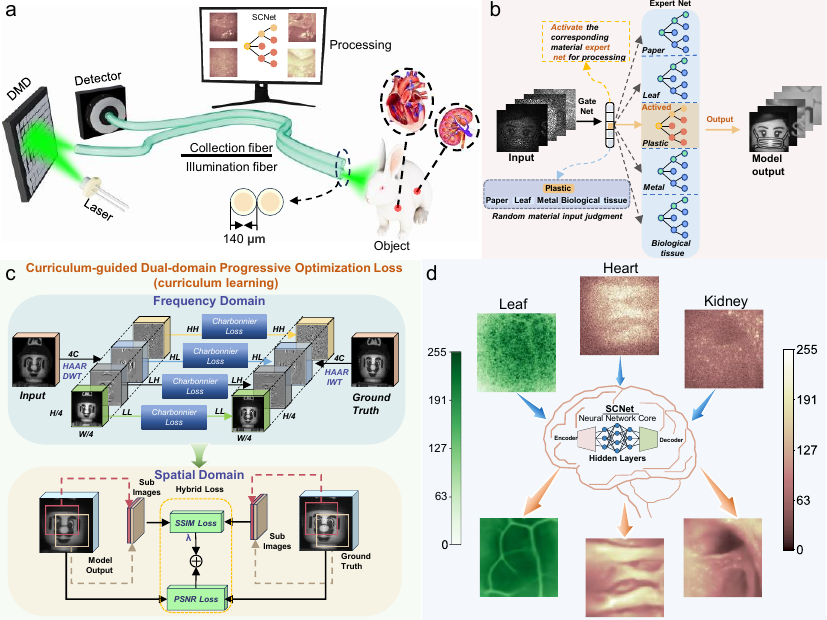}
\refstepcounter{figure}
\caption*{\textbf{Fig.1}\textbar\textbf{SCNet speckle suppression model architecture for ultrathin multimode-fiber endoscopic imaging.}
\textbf{a,} Schematic of the holographic endoscopic system. A DMD shapes the wavefront to focus light through an illumination MMF onto a distant object. Backscattered signal is collected via a second parallel MMF to a bucket detector. 
\textbf{b,} Architecture of the SCNet model. The network features a MoE block controlled by a gating mechanism that dynamically weights material-specific experts to handle diverse scattering properties. 
\textbf{c,} Dual-domain curriculum learning strategy. The model is trained first by optimizing for frequency domain consistency using wavelet transforms to learn noise-invariant features, followed by spatial domain fine tuning to recover high-frequency morphological details. 
\textbf{d,} Representative SCNet reconstructions for a leaf and mammalian tissues (rabbit heart and kidney), showing the raw speckled images (top) and the corresponding SCNet outputs (bottom).}\label{fig1}
\end{figure}

We built an ultrathin endoscopic imaging system for distant objects observation using two multimode fibers (\SI{100}{\micro m} core, \SI{140}{\micro m} cladding), with one fiber delivering holographically shaped illumination and the second fiber collecting backscattered photons to a bucket detector (Fig.\ref{fig1}a). 
By employing wavefront shaping via a digital micromirror device (DMD), we generated diffraction-limited foci in the far field of the distal facet to raster-scan targets. 
In this ultrathin regime, the reconstructed images are dominated by high-contrast speckle fluctuations that obscured fine structural details.

To reconstruct high-fidelity images from these speckle-degraded measurements, we developed SCNet, a speckle suppression model designed for ultrathin MMF endoscopy. SCNet uses a MoE architecture in which a lightweight gating network predicts an input regime and routes the measurement to a corresponding expert branch (Fig.\ref{fig1}b). 
In our implementation, the expert set is specialized for distinct material categories, and the gating network performs the routing automatically without manual selection. The full network definition is shown in Extended Data Fig.\ref{fige4}a.

We trained SCNet using a dual-domain curriculum learning strategy that combines frequency domain supervision with spatial domain fidelity constraints (Fig.\ref{fig1}c). In the frequency domain, inputs and targets are decomposed with a wavelet transform and optimized with multi-stage losses on sub-bands to progressively enforce consistency from low to high frequency components. 
In the spatial domain, SCNet is optimized using a hybrid objective that promotes structural similarity and pixelwise fidelity on sub-images (Fig.\ref{fig1}c). This dual-domain schedule is used throughout training to emphasize morphology consistent restoration while limiting sensitivity to specific speckle realizations.

Representative processing results across distinct sample types are shown in Fig.\ref{fig1}d Compared with the raw speckled images (Fig.\ref{fig1}d, top), SCNet outputs (Fig.\ref{fig1}d, bottom) recover continuous leaf venation patterns and enhance fine anatomical texture in mammalian organs (heart and kidney) under the same single fiber collection setting.

\subsection{Depth-invariant 3D object reconstruction}\label{subsec2}
\begin{figure}[htbp!]
\centering
\includegraphics[width=1\textwidth]{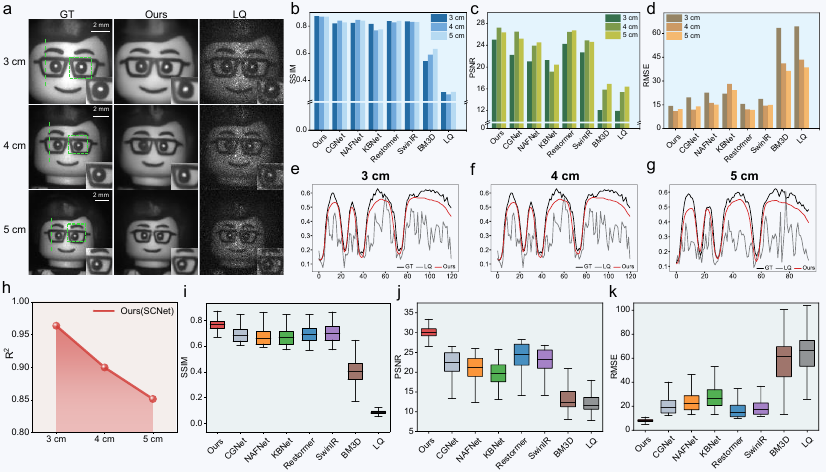}
\refstepcounter{figure}
\caption*{\textbf{Fig.2}\textbar\textbf{Depth-invariant reconstruction of 3D polymeric phantoms via dynamic expert routing.}
\textbf{a,} Comparison of raw and processed images of Lego minifigures at different working distances (3, 4, 5 cm). Raw images (LQ) were captured using a single-core ultrafine multimode fiber with a core diameter of 100 µm; processed images (Ours) represent results obtained via the SCNet model; ground truth images (GT) were acquired using a fiber bundle comprising three fibers with a core diameter of \SI{1000}{\micro m}. The enlarged region in the lower right corner of the original image demonstrates the detail recovery in the eye and eyeglass areas.
\textbf{b-d,} Quantitative comparison of different models across three metrics-Structural Similarity Index (SSIM), Peak Signal-to-Noise Ratio (PSNR), and Root Mean Square Error (RMSE) at each working distance shown in (a). Higher SSIM and PSNR values indicate better performance, while lower RMSE values denote superior performance.
\textbf{e-g,} Intensity distribution curves extracted along the facial region (green dashed line) at working distances of 3 cm (e), 4 cm (f), and 5 cm (g). Black, gray, and red curves represent intensity distributions for GT, LQ, and SCNet speckle-removed output, respectively.
\textbf{h,} Based on the intensity distributions in (e-g), the coefficient of determination (R²) quantifies the goodness of fit between SCNet outputs and GT.
\textbf{i-k,} Overall performance comparison of different models on the full plastic sample test set, presented as box plots for SSIM (i), PSNR (j), and RMSE (k). Box plots show the median (box line), interquartile range (box width), and 1.5 times the interquartile range (whiskers).}\label{fig2}
\end{figure}

To validate the model’s capacity to handle scattering variations induced by complex 3D geometries, we imaged polymeric phantoms (Lego minifigures) at working distances ranging from 3 cm to 5 cm (Fig.\ref{fig2}). 
In lensless MMF far-field endoscopy, shifting the imaging plane alters the effective field-of-view and resolution, a phenomenon that typically causes fixed-weight networks to fail when object topography varies. 
To establish a rigorous ground truth (GT) for MMF far-field imaging measurements, we utilized a high collection efficiency probe comprising three large-core multimode fibers (core diameter: 1000 µm, NA: 0.37), while the test input (LQ) was restricted to a single ultrathin collection fiber (core diameter: 100 µm, NA: 0.22) to mimic the minimally invasive constraint (Fig.\ref{fig2}a).

We challenged SCNet with these axial variations by raster-scanning the phantoms across the 2 cm depth range. While the raw single-fiber reconstructions exhibited severe depth-dependent speckle fluctuations that obliterated facial features and surface textures (Fig.\ref{fig2}a, LQ), SCNet consistently recovered high-fidelity structural details across the full depth of field. 
This robustness stems from the multi-scale nature of the network's frequency encoding. Unlike standard convolutional layers that operate on a fixed spatial scale, the integrated discrete wavelet transforms (DWT) decompose the input into distinct frequency sub-bands. 
This allows the model to separate the fundamental morphological structure from the distance-dependent variation in speckle grain size, ensuring stable reconstruction without distance-specific recalibration.

Quantitative benchmarking against state-of-the-art restoration models, including CGNet\cite{Ghasemabadi_2024}, NAFNet\cite{chen2022simple}, KBNet\cite{zhang2023kbnet1}, Restormer\cite{zamir2022restormer}, SwinIR\cite{liang2021swinir}, and BM3D\cite{danielyan2011bm3d}, demonstrated the superior generalization of our physics-guided approach. SCNet consistently achieved the highest structural similarity (SSIM $\sim{0.85}$) and peak signal-to-noise ratios (PSNR\textgreater26 dB) across all working distances (Fig.\ref{fig2}b-d). 
In contrast, conventional deep learning models failed to differentiate signal from non-stationary speckle patterns, resulting in significant artifacts and lower fidelity scores (Fig.\ref{fig2}i-k). Line profile analysis further confirmed that SCNet maintains high fidelity in intensity distribution ($R^2 > 0.95$ at 3 cm), validating its ability to model the deterministic input-output mapping of MMFs even under varying propagation conditions (Fig.\ref{fig2}e-h). 
Detailed comparisons of edge preservation and texture recovery against other baseline models are provided in Extended Data Fig.\ref{fige1}.

\subsection{Resolving macroscopic features on fibrous media}\label{subsec3}

\begin{figure}[htbp!]
\centering
\includegraphics[width=1\textwidth]{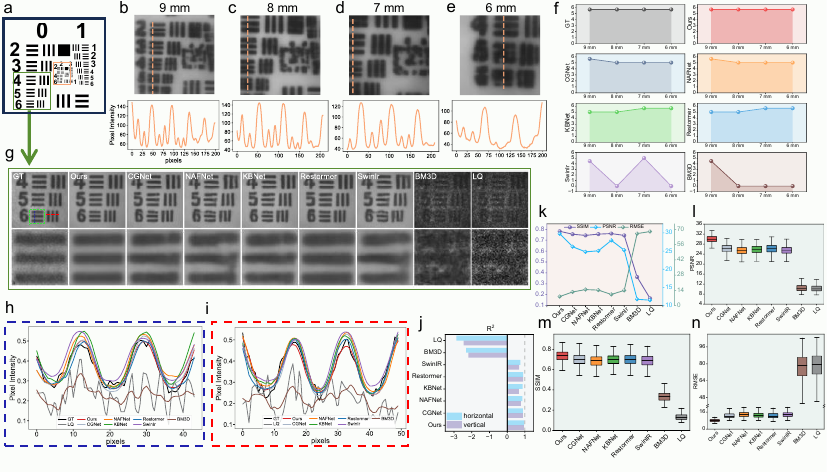}
\refstepcounter{figure}
\caption*{\textbf{Fig.3}\textbar\textbf{High-resolution imaging on absorbent paper via spectral consistency learning.}
\textbf{a,} Schematic of the standard USAF-1951 resolution target. 
\textbf{b-e,} Reconstruction of Group 2-3 Elements at working distances of 9, 8, 7, and 6 mm. 
Top row: SCNet outputs showing resolved line pairs. Bottom row: Corresponding vertical line sections (yellow dashed lines), used to determine the maximum resolvable spatial frequency based on peak-to-valley contrast. 
\textbf{f,} Maximum resolvable spatial frequency (line pairs/mm) achieved by SCNet compared to baseline models (CGNet, NAFNet, KBNet, Restormer, SwinIR, and BM3D) across all distances. SCNet consistently resolves 5.66 lp/mm (Group 2, Element 4). 
\textbf{g,} Macroscopic reconstruction of Group 0 Elements 4-6 from the USAF target; green dashed box marks the zoomed in region. 
\textbf{h,i,} Line-pair intensity profiles along the Element 6 horizontal and vertical lines (blue and red dashed lines in \textbf{g}) for all methods, including GT and LQ. 
\textbf{j,} Comparison of the goodness-of-fit ($R^2$) between each method and the GT profiles in \textbf{h} and \textbf{i}. 
\textbf{k,} Quantitative evaluation of speckle reduction performance across models on the images shown in \textbf{g}, including SSIM, PSNR, and RMSE metrics.
\textbf{l-n,} Overall performance comparison on the entire paper test set, presented as box plots showing SSIM (l), PSNR (m), and RMSE (n).}\label{fig3}
\end{figure}

To quantify the recoverable spatial resolution on a highly absorbent, fibrous substrate, we printed a USAF-1951 resolution target on paper and imaged it using the same MMF endoscope configuration used throughout this study(Fig.\ref{fig3}a). 
Unlike the dielectric surface of plastic (Fig.\ref{fig2}), paper introduces significant volume scattering and photon diffusion, which severely degrades the modulation transfer function of the backscattered signal. 
In this regime, conventional end-to-end models trained with pixel-wise losses (e.g., Mean Squared Error) typically suffer from spectral bias, yielding over-smoothed reconstructions that obliterate high-frequency details \cite{tancik2020fourier}.

SCNet circumvents this limitation through its dual-domain curriculum learning strategy (Curriculum-guided Dual-domain Progressive Optimization Loss, CDPO-Loss). 
During the initial training phase, we prioritize the restoration of high-frequency wavelet subbands (LH, HL, HH) by assigning distinct weights to the four subbands derived from the Haar wavelet transform within the loss function. This strategy enables the model to explicitly learn to discriminate morphological edges from speckle noise prior to fine-tuning for spatial consistency.
Consequently, SCNet consistently resolved line patterns up to Group 2, Element 4 (5.66 line pairs/mm) across working distances of 6-9 mm (Fig.\ref{fig3}b-e), effectively matching the ground truth resolution acquired by the large-core fiber bundle (Fig.\ref{fig3}f and Extended Data Fig.\ref{fige2}).

Comparative analysis reveals that state-of-the-art restoration models (e.g., SwinIR, CGNet) failed to resolve these frequencies, succumbing to noise-induced blurring or hallucinating artifacts (Fig.\ref{fig3}f). 
We further validated this capability on the macroscopic elements of the same target (Group 0, Elements 4-6), where SCNet maintained superior contrast-to-noise ratios and accurate edge definition compared to the raw input (Fig.\ref{fig3}g). 
Line-pair intensity profiles extracted along orthogonal directions (Fig.\ref{fig3}h,i) and their goodness-of-fit to GT ($R^2$; Fig.\ref{fig3}j) showed close agreement between SCNet outputs and the GT profiles. 
Image-quality metrics for the example in Fig.\ref{fig3}g (SSIM, PSNR and RMSE; Fig.\ref{fig3}k) and quantitative benchmarking across the full paper test set confirmed that this spectral consistency translates to statistically significant improvements in SSIM ($\sim{0.8}$), PSNR (\textgreater 28 dB) and RMSE (\textless 0.2) compared to fixed-weight baselines (Fig.\ref{fig3}l-n and Extended Data Fig.\ref{fige2}), establishing the framework's ability to recover high-fidelity information from highly scattering, non-cooperative surfaces.
Additional paper target evaluations are provided in the Supplementary Information, including systematic comparisons of CDPO-Loss versus MSE-Loss and Contrast Limited Adaptive Histogram Equalization (CLAHE) post-processing across working distances (Supplementary Information S6), and traditional filtering baselines on the paper USAF target (Supplementary Fig.19). 
The dual-domain training objective that underpins this paper resolution benchmark is described in Methods, where wavelet-domain supervision is used before spatial-domain fine tuning.

\subsection{Universal generalization across diverse scattering regimes}\label{subsec4}

\begin{figure}[htbp!]
\centering
\includegraphics[width=1\textwidth]{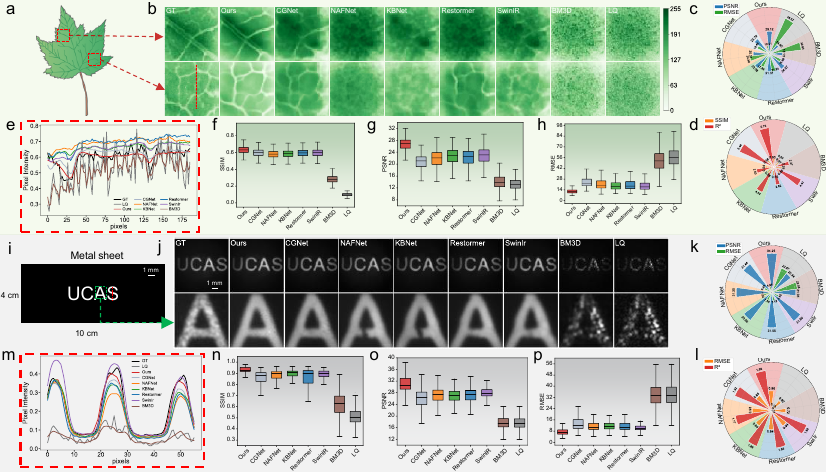}
\refstepcounter{figure}
\caption*{\textbf{Fig.4}\textbar\textbf{Complex structure restoration across biological and metal surfaces.}
\textbf{a,} Schematic of the leaf sample, with the red box indicating the imaging acquisition area.  
\textbf{b,} Reconstruction of leaf vasculature.
\textbf{c,d,} Quantitative performance (PSNR, RMSE, SSIM, $R^2$) on leaf samples.
\textbf{e,} Line intensity profiles across a secondary vein. 
\textbf{f-h}, Statistical performance distributions (SSIM, PSNR, RMSE) for the complete leaf test set.
\textbf{i,} Schematic of the metallic sample showing engraved characters, with the green dashed box marking the magnified “A” region and the red dashed line showing the intensity extraction path along the letter “S”. 
\textbf{j,} Metal surface speckle removal. Top row shows full-field reconstructions; bottom row shows magnified views of the “A” region for each method. \textbf{k,l,} Quantitative performance (PSNR, RMSE, SSIM, $R^2$) on metal samples. \textbf{m,} Line intensity profile across the sharp edge of the letter “S”. \textbf{n-p,} Statistical performance distributions for the complete metal test set.
}\label{fig4}
\end{figure}

To validate SCNet as a generalized foundation model, we challenged it with two optically distinct substrates, semi-transparent plant leaves and opaque metallic components (Fig.\ref{fig4}). 
This experiment directly addresses the domain shift caused by material-dependent light transport, in which biological tissues exhibit volume scattering with diffusive speckle while metallic surfaces produce specular reflections with sharp macroscopic edges.

For the biological plant sample, we imaged maple leaves, where the internal vasculature is typically obscured by multiple scattering events within the mesophyll layer (schematic as Fig.\ref{fig4}a). While the raw backscattered input severely blurred the fine secondary venation (Fig.\ref{fig4}b, LQ), SCNet successfully reconstructed the complete vascular network, resolving both the thick primary veins and the intricate capillary-like secondary veins (Fig.\ref{fig4}b). 
Quantitatively, we focused on the more demanding lateral vein region. Radial bar plots of PSNR and RMSE (Fig.\ref{fig4}c) and of SSIM and $R^2$ (Fig.\ref{fig4}e) derived from line profiles (Fig.\ref{fig4}d) showed that SCNet achieved the best scores among all comparison methods, with intensity profiles closely matching the GT in this region. Evaluation over the full leaf test set confirmed that SCNet maintained a consistent advantage in SSIM, PSNR and RMSE relative to deep-learning baselines (Fig.\ref{fig4}f-h). 
Traditional filtering-based approaches for leaf speckle suppression were also assessed in the Supplementary Information and were found to neither remove coherent noise nor recover the underlying vein anatomy (Supplementary Fig.21). The system’s MoE architecture was critical here. The gating network identified the diffusive scattering signature and routed the signal to experts specialized in low-frequency structural recovery, achieving high consistency with GT intensity profiles.
This gating network is implemented as an Enhanced Cascaded Classifier Network (ECCNet); details of its architecture are provided in Supplementary Information S8.

We next tested SCNet on a stainless steel sheet surface patterned with “UCAS” letters, corresponding to the acronym of the University of Chinese Academy of Sciences (Fig.\ref{fig4}i,j). In the raw single-fiber reconstructions, letter boundaries were strongly corrupted by speckle, whereas SCNet produced clear letter shapes and sharp corners in both the full field and the magnified “A” region (Fig.\ref{fig4}j). For the metal sample, the challenge shifted to preserving sharp edges against specular glare. Here, the gating mechanism automatically activated experts trained on surface-scattering priors, allowing SCNet to recover the crisp boundaries of the characters. 
Results showed that SCNet achieved the highest PSNR and SSIM and the lowest RMSE on this metal sample (Fig.\ref{fig4}k,l). Line intensity profiles along the S-shaped path (Fig.\ref{fig4}m) revealed that SCNet closely tracked the GT curve, with the highest $R^2$ among all methods (Fig.\ref{fig4}l). Statistics across the metal test set further indicated stable gains in SSIM, PSNR and RMSE (Fig.\ref{fig4}n-p). Supplementary Fig. 8 provides additional comparisons of loss configurations (MSE-Loss, CDPO-Loss and CLAHE) and shows that CDPO-Loss based SCNet yields the best trade off between edge integrity, structural detail and artifact suppression on metal surfaces.

\subsection{Organ imaging and acceleration}\label{subsec5}

\begin{figure}[htbp!]
\centering
\includegraphics[width=1\textwidth]{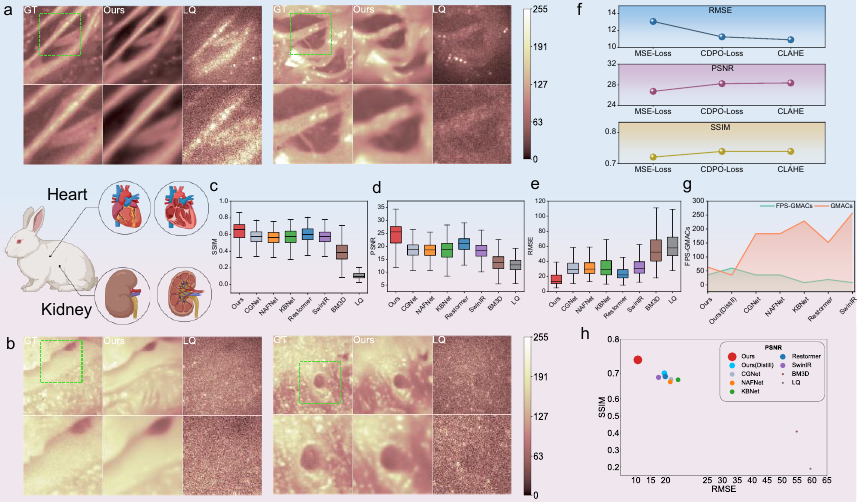}
\refstepcounter{figure}
\caption*{\textbf{Fig.5}\textbar\textbf{Biological organ imaging and computational optimization.} 
\textbf{a,b,} Representative reconstructions of rabbit heart (a) and kidney (b) tissues, shown as GT, SCNet output and LQ; the second row shows magnified views of the green dashed regions.
\textbf{c-e,} Statistical performance comparison of SCNet with competing methods for biological test sets, showing as SSIM (c), PSNR (d), and RMSE (e).
\textbf{f,} Comparison of image processing performance metrics for MSE loss function, CDPO loss function, and CLAHE histogram equalization, shown as RMSE, PSNR and SSIM. 
\textbf{g,} Comparison of computational complexity (GMACs) and inference speed (FPS) across all models.
\textbf{h,} Average performance summary across five material test sets; vertical axis shows average SSIM, horizontal axis shows average RMSE and bubble size indicates average PSNR.}\label{fig5}
\end{figure}

While phantom studies validate physical robustness, clinical endoscopy demands the resolution of low contrast anatomical features within dynamic scattering biological environments. 
To evaluate SCNet in this physiological regime, we imaged isolated rabbit heart and kidney tissues (Fig.\ref{fig5}a,b). 
Unlike synthetic targets, soft tissues exhibit subtle refractive index fluctuations that result in low-contrast speckles, often causing standard restoration models to fail in preserving fine texture. 
To overcome this, we integrated CLAHE into the training pipeline, effectively amplifying high-frequency tissue signatures during the learning process.

In the raw reconstructions, myocardial fiber texture in the heart and fine renal structures in the kidney were obscured by speckle, whereas SCNet recovered coherent tissue morphology in both organs, as highlighted in the magnified regions (Fig.\ref{fig5}a,b). 
After SCNet processing, heart images revealed continuous muscle fascicles and major vessel lumens, while kidney images recovered the tubular structures in the cortex and the complex anatomy of the pelvis and calyces that are barely discernible in the raw reconstructions (Fig.\ref{fig5}a,b and Extended Data Fig.\ref{fige3}).

Across the heart and kidney full test sets, SCNet achieved the best overall performance among all comparison methods, including CGNet, NAFNet, KBNet, Restormer, SwinIR and BM3D, with consistent gains in SSIM and PSNR and reduced RMSE (Fig.\ref{fig5}c-e). 
Additional qualitative comparisons across competing baselines for the same tissue positions are provided in Extended Data Fig.\ref{fige3} and Supplementary Fig.2.

To disentangle the contributions of our optimization strategy and contrast enhancement, we compared three SCNet variants: a baseline trained with a standard MSE-loss, a model trained with the CDPO-Loss, and a CLAHE as a post-processing step (Fig.\ref{fig5}f). 
It shows that CDPO-Loss consistently improved SSIM and PSNR relative to the MSE baseline while reducing RMSE  and such full tests comparison across all five material datasets is in Supplementary Fig.5. 
Applying CLAHE on top of the CDPO-Loss model yielded a further gain in image quality metrics and produced more uniform local contrast, particularly in low-contrast regions of soft tissue, while preserving the structural details already recovered by SCNet (Supplementary Figs.6-10). 
As summarized in Extended Data Table 1, the CDPO-Loss and CLAHE configurations jointly define the highest performing variant of the framework.

We also optimized the model for deployment under the latency and power constraints of clinical endoscopy.
Using a multi-teacher knowledge-distillation scheme, we compressed the full SCNet-MoE architecture into a lightweight student network, SCNet-Distill, that retains the physics-guided DWT/IDWT(Haar Inverse Discrete Wavelet Transform) blocks but uses fewer layers and reduced channel width (Methods and Extended Data Fig.\ref{fige4}b). 
This compression lowered the computational cost by 55\% from 63.43 to 34.25 GMACs (Giga Multiply-Accumulate Operations, GMACs) (Fig.\ref{fig5}g) while maintaining reconstruction quality close to the teacher model (PSNR reduction \textless 4 dB on biological tissues, Extended Data Table 1,2; Supplementary Information S9). 
Inference tests show that SCNet-Distill reaches 60 FPS, approximately doubling the frame rate of the teacher model and exceeding all general-purpose baselines, which operate at \textless36 FPS with substantially higher GMACs (Fig.\ref{fig5}g and Extended Data Table 2). 
Aggregated over all five material test sets, SCNet variants occupy the favourable region of the performance and efficiency trade-off, with higher SSIM, lower RMSE and competitive PSNR relative to competing methods (Fig.\ref{fig5}h).

\section{Discussion}\label{sec3}

We have presented SCNet, a physics-guided foundation model that effectively resolves the longstanding trade-off between miniaturization and image quality in ultrathin multimode fiber far-field imaging. By enabling high-fidelity imaging through a \SI{100}{\micro m} collection aperture, our method transforms the dominant limitation of ultrathin probes, namely speckle arising from severely photon-limited single-fiber collection, into a tractable computational reconstruction problem. 
Unlike previous learning-based approaches that are restricted to specific samples or rely on extensive physical recalibration, SCNet demonstrates universal applicability across diverse industrial and biological scattering regimes without retraining. 
This establishes a new paradigm for an optical endoscopy foundation model capable of dynamically suppressing complex backscattering interference to restore clinically valuable sub-millimeter structures in real time.

The success of SCNet stems from the integration of physical priors into deep neural computation, marking a methodological transition from simply fitting noise distributions to deconstructing the physical generation process of the image. Traditional end-to-end paradigms often struggle with the multiplicative, non-stationary nature of speckle noise that leading to over smoothing artifacts. 
By incorporating a discrete wavelet transform module, SCNet explicitly disentangles structural information (low-frequency sub-bands) from noise dominated components (high-frequency sub-bands) in the frequency domain. 
This design acts as a rule-based spatial attention mechanism, guiding the network to distinguish reducible fluctuations from essential structure details. 
Furthermore, the adoption of a MoE architecture addresses the challenge of diverse material generalization. Different materials impose distinct scattering signatures on the backscattered light field, ranging from volume scattering in tissues to specular reflections on metal. 
While monolithic models often suffer from catastrophic forgetting when trained on such heterogeneous datasets, SCNet's adaptive gating mechanism dynamically allocates computational resources to the most relevant expert network. 
This ensures that the model captures material-specific statistical invariants while maintaining a unified inference framework.

Bridging the gap between algorithmic performance and clinical utility requires addressing both convergence stability and computational latency. Our dual-domain curriculum learning strategy mirrors the logic of human expert systems by optimizing for frequency-domain consistency before fine-tuning for spatial fidelity metrics, the model avoids the local minima associated with pixel-wise loss functions. 
When combined with CLAHE-based post-processing, this approach effectively amplifies high-frequency structure signatures that are otherwise lost in low-contrast environments. Crucially, we translated this high fidelity restoration into a clinically viable tool through model distillation. 
By transferring knowledge from the teacher models (comprising the MoE architecture) to a single lightweight student network via distillation, we reduced the computational cost by 55\% while maintaining image quality, achieving an inference speed of 60 FPS (Extended Data Table 2). 
This performance sets image reconstruction refresh rate only by the point scanning speed of MMF imaging system , establishing SCNet as a feasible platform for dynamic imaging contexts.

Despite these advancements, limitations remain in the current implementation. Our system employs a dual-fiber configuration (combined diameter \SI{280}{\micro m}) to separate illumination from collection, a design necessitated by the need to mitigate background noise from tip reflections. 
While this footprint is ultrathin compared to conventional endoscopes, the ultimate goal of minimally invasive access requires single-fiber operation. 
Future iterations will target this by integrating advanced distal micro-optics or coatings to suppress interface reflections, potentially halving the probe dimension. 
On the computational side, the gating network currently relies on pre-trained classifiers, which introduces a disjointed optimization process. 
Future work could explore end-to-end differentiable routing mechanisms to further enhance adaptability to unseen scattering media. Additionally, while SCNet demonstrates robustness on isolated tissues, rigorous validation in large animal models is required to standardize performance in highly dynamic, fluid-filled physiological cavities. 
Looking forward, the scalable framework of SCNet can be readily adapted to other computational imaging modalities governed by differential forward models, such as optical coherence tomography\cite{lee2017deep} and photoacoustic imaging\cite{grohl2021deep}, opening new avenues for precise, real-time deep tissue diagnostics.


\section{Methods}\label{sec4}
\subsection{Experimental setup}\label{subsec12}

We implemented a lensless, dual multimode fiber holographic endoscope that forms reflectance images by measuring the local response to a scanned diffraction-limited focal spot. 
A continuous-wave laser is wavefront-shaped by a DMD and coupled into an illumination , while the backscattered signal is collected by a second, parallel MMF and routed to a high-sensitivity detector (Fig.\ref{fig1}a). 
Separating illumination and collection improves signal recovery at stand-off working distances and reduces background contributions from Fresnel reflections at the fiber facets. 
Because only a limited angular portion of the scattered field is accepted by the fiber core, the detected intensity at each scan coordinate primarily reflects local reflectivity and surface-dependent scattering, including roughness, orientation and axial depth variations. 
Under the ultrathin collection constraint, this restricted acceptance produces strongly speckle-dominated measurements; when a high-SNR reference or ground truth is required, we additionally use three large-core collection fibers to increase the acceptance area and spatially average speckle.

A detailed optical layout is provided in Supplementary Fig.1. Briefly, a 532-nm laser source is split into signal and reference arms. 
In the signal arm, the beam is expanded to illuminate the DMD at an incidence angle of $\sim{24}^{\circ}$. 
We operate the DMD in an off-axis holographic configuration to approximate phase-only modulation using a Lee hologram scheme, and two 4f relay optics select the first diffraction order, scale the field to match the MMF core and image the modulated wavefront onto the proximal facet. 
The DMD encodes two symmetric regions in the Fourier plane that carry orthogonal linear polarizations, which are recombined into a common propagation path using a polarization beam displacer before coupling into the illumination fiber.

To calibrate and focus through the MMF, the distal output field is relayed to a camera and interfered with the reference arm to record off-axis holograms. We recover the complex field using phase-shifting interferometry and measure the fiber transmission matrix (TM), which captures the linear mapping from input modes to distal outputs for a fixed probe configuration. 
The TM is measured using an in situ wavefront-correction strategy that compensates static aberrations across the optical path and yields diffraction-limited foci at the calibration plane \cite{vcivzmar2010situ}. We use an orthogonal plane-wave input basis and define a Cartesian grid of diffraction-limited focal spots as the output basis. 
From the measured TM, we compute optimized binary DMD patterns that generate far-field foci and raster scan them across the target to form reflectance images.

Maintaining TM validity requires that the optical setup remain stationary between calibration and imaging and that the relay optics preserve pupil geometry. 
To limit drift-induced phase errors that would broaden the focus and reduce image contrast, we used low-drift optomechanics, placed the setup on an actively damped optical table and operated in a temperature-controlled environment. 
Under these conditions, imaging performance was stable for several hours without recalibration. 
In practice, we repeat calibration when a new MMF segment or probe is installed, following standard TM acquisition protocols \cite{ploschner2015seeing_m}. 
Using a MEMS-based DMD enables rapid focus refreshing at tens of kilohertz, supporting high-speed raster scanning. 
In this study, we measured TMs with 7,500 input modes and up to 124,870 output modes, where the latter was set by the desired field of view. 
TM acquisition and pattern synthesis were performed on a workstation equipped with an Intel i9-14900K CPU (128 GB RAM) and an NVIDIA RTX 4070Ti Super GPU.

\subsection{Dataset construction}\label{subsec13}

All the data were simultaneously obtained with high-quality ground truth and low-quality speckle images during the collection process. The original image size was 416$\times$416 pixels, which was then uniformly cropped to 220$\times$220 pixels for model training and evaluation. 
The imaging objects include plastic, paper, metal, plant leaves, rabbit heart and kidney. 
The dataset was independently constructed according to material categories, as follows:

Plastic: Use Lego figurines. By altering their poses, accessories, and imaging distance (1-5 cm), diversity is enhanced. 
The figurines are divided into training set (9,000 pairs of images), validation set (900 pairs of images), and test set (900 pairs of images). 
The test set only contains frontal, complete facial images for final presentation and evaluation.

Paper: Use A4-sized paper printed with the USAF-1951 resolution target and other English texts in Arial 10 pt font. 
The paper contains a total of 30,000 pairs of training images, 3,000 pairs of validation images, and 3,000 pairs of test images. 

Metal: Made with custom stainless steel sheets (dimension of 10$\times$4 cm), engraved with various text. 
It consists of 20,000 pairs for the training set, 2,000 pairs for the validation set, and 2,000 pairs for the test set of images. 

Leaf: Utilized various plant leaves including maple leaves, cherry blossom leaves and euphorbia leaves. 
The dataset consists of 30,000 pairs for training, 3,000 pairs for validation, and 3,000 pairs for testing.

Biological tissue: Utilizes organs such as the rabbit heart and kidneys. 
The dataset consists of a training set of 70,000 pairs, a validation set of 7,000 pairs, and a test set of 7,000 pairs of images.

For samples with high reflectivity and macroscopic structures such as plastic, paper and metal, the imaging distance range is set as 1-5 cm. 
For biological samples with low reflectivity and complex microscopic structures such as leaves, hearts and kidneys, the imaging distance range is 0.1-3 cm. 
To expand the data set and improve the generalization performance of the model, we combined the moving cropping and random cropping methods for the original image. 
The image values were normalized to a range of 0-1. 
Through experiments, it was found that when the sequential cropping stride is 60 and the number of random cropped images is 6, the model trained has the best speckle elimination performance and generalization performance. 
The data set's augmentation strategy for cropping is based on the standard practice in the field of super-resolution imaging by using fixed stride cropping and random cropping to expand data diversity, while avoiding the loss of structural information caused by excessive cropping \cite{Fang_2021_03,Chen_2023_05}.

\subsection{Animal}\label{subsec14}

All animal experiments were conducted in accordance with the protocols approved by the relevant animal ethics and usage committees. Twenty SPF-grade New Zealand rabbits (half male and half female, with a body weight of 2.0-2.5 kg) were selected and acclimated for one week before the operation.

After the animals were deeply anesthetized with isoflurane (maintained at a concentration of 2-3\%), they were euthanized by bloodletting from the carotid artery. Subsequently, the abdominal aorta was perfused, and the vessel was rinsed with normal saline until the effluent was colorless. Then, it was perfused with 4\% paraformaldehyde until the heart stopped beating and the organs became hardened.

After extracting the heart tissue, the surface fat was removed, and a vertical incision was made along the long axis of the left ventricle to completely expose the atrium, ventricle cavities and internal structures. After extracting the renal tissue, it was half-cut along the coronal plane, clearly showing the renal cortex, medulla and renal pelvis. 
All the tissue samples were placed in petri dishes lined with sterile wet gauze after being trimmed, and the subsequent imaging operations were carried out in a laminar flow hood to ensure that the sections were neat and the structures were clear.
The processing procedure for animal tissues follows the standard protocols in the field of biological imaging: 4\% paraformaldehyde perfusion fixation is based on the EM imaging processing method for rat hippocampal tissue \cite{renier2014idisco}, which can effectively preserve the microscopic structures of heart muscle fibers and renal tubules; while the trimming method for the longitudinal section of the left ventricle of the heart and the coronal section of the kidney is consistent with the fine operation logic for exposing the synaptic structures of neurons, ensuring the structural integrity within the imaging field \cite{mishchenko2010ultrastructural}.

\subsection{Hybrid expert architecture}\label{subsec15}

MoE architecture forms the core component of SCNet, enabling adaptive processing of speckle patterns from diverse material types. 
As illustrated in Fig.\ref{fig1}b, the MoE framework comprises two fundamental elements: a gating network and multiple expert networks.

The gating network operates as a lightweight material classifier, pre-trained to categorize input speckle images into five distinct material classes: plastic, paper, metal, plant leaves, and biological tissues. 
This classification is based on the unique scattering characteristics exhibited by each material type when imaged through ultrathin multimode fibers. The gating mechanism employs a sparsely-activated routing strategy, where only the most relevant expert network is activated for each input sample, thereby optimizing computational efficiency while maintaining specialized processing capabilities.

Each expert network within the MoE framework is constructed upon the wavelet U-Net architecture described in Section \ref{subsec16}, but independently trained on material-specific datasets. 
This specialized training allows each expert to develop deep domain knowledge for handling the particular scattering properties and noise characteristics of its assigned material category.
For instance, the metal expert network learns to preserve sharp edge features commonly found in metallic samples, while the biological tissue expert focuses on reconstructing fine anatomical structures such as vascular networks and cellular arrangements.

The routing mechanism employs a softmax-based attention strategy to compute the probability distribution over experts:
\begin{equation}
P_i = \frac{\exp(W_i \cdot x + b_i)}{\sum_{j=1}^N \exp(W_j \cdot x + b_j)}
\end{equation}
where $P_i$ represents the routing probability for expert $i$, $x$ denotes the feature representation from the gating network, and $W_i$, $b_i$ are the corresponding weight and bias parameters. Only the expert with the highest probability receives the input for processing, while others remain inactive, ensuring computational sparsity.

The MoE framework enables SCNet to maintain high reconstruction fidelity across diverse materials without requiring material-specific retraining or manual configuration, establishing it as a truly universal foundation model for ultra-thin fiber endoscopic imaging.

The implementation of the MoE architecture follows the sparse activation paradigm established in large-scale neural networks, where conditional computation enables efficient scaling to multiple domains \cite{guo2018multi, li2022sparse, jiang2024medmoe}.  
This approach has been particularly effective in medical imaging applications requiring adaptation to diverse tissue types, demonstrating robust performance across varying imaging conditions and sample characteristics \cite{ding2025denseformer, wang2025test, luo2025rethinking, chen2024multi}.

\subsection{Network architecture}\label{subsec16}

The SCNet model is overall based on the U-Net architecture, as shown in Fig.\ref{fig1}. Its core innovation lies in the introduction of wavelet transformation and a hybrid expert model.
The U-Net architecture of SCNet is based on the cascaded convolution design of U-Net \cite{Ronneberger_2015, Ghasemabadi_2024}, which has been proven to be capable of efficiently extracting multi-scale imaging features.

Wavelet domain processing. A Haar discrete wavelet transform layer is placed at the input end of the encoder to decompose the input image into four frequency subbands: LL, LH, HL, and HH. The spatial dimension of the feature map is reduced to 1/4 of the original image. 
In the output end of the decoder, inverse wavelet transform is used to reconstruct the spatial image.
The introduction of wavelet transform (Haar discrete wavelet decomposition) is based on the research conclusion of multi-scale structural similarity - frequency domain decomposition can effectively decouple noise and structural information \cite{Xu_2023_11}; 
The gated routing mechanism of the hybrid expert model is consistent with the Pareto optimization idea of “task classification - dedicated network” in multi-task learning, which can avoid the performance degradation of a single network when adapting to multiple materials \cite{Sener_2018_10}.
Enhanced Jump Connection. The standard jump connection is replaced with a learnable module that includes convolutional layers and the GELU activation function, enabling adaptive extraction of multi-scale features.

Hybrid expert architecture. SCNet employs a two-stage process Fig.\ref{fig1}b. Firstly, a pre-trained material classification network acts as a gating network to classify the input speckle images (plastic, paper, metal, leaf, biological tissue). 
Subsequently, the images are routed to the dedicated expert networks for speckle removal corresponding to each material. Each expert network is independently pre-trained based on the aforementioned wavelet U-Net architecture. 
The detailed network structure is shown in Extended Data Fig.\ref{fige4}b.

To overcome the inherent limitations of spatial-domain convolutions in distinguishing multiplicative speckle noise from high-frequency tissue details, we introduce a frequency-aware architecture, Haar wavelet. 
Unlike traditional approaches that operate solely on pixel intensities, our method leverages the DWT to explicitly decouple structural information from noise components prior to feature extraction.

The integration of the DWT into the SCNet architecture introduces a fundamental shift in the signal processing paradigm. 
Unlike standard spatial-domain convolutional networks that must implicitly learn to decouple high-frequency noise from low-frequency structural components.

Laser speckle noise is characterized by high-frequency granular patterns that act multiplicatively on the coherent signal. 
In a standard Convolutional Neural Network (CNN), the filters in the initial layers must simultaneously perform feature extraction and noise filtering, a dual objective that often leads to optimization difficulties known as the “spectral bias” of neural networks.

The DWT module in the proposed architecture mathematically transforms the input tensor $\mathbf{X} \in \mathbb{R}^{H \times W}$ into four orthogonal sub-bands:
\begin{equation}
\text{DWT}(\mathbf{X}) = [\mathbf{X}_{LL}, \mathbf{X}_{LH}, \mathbf{X}_{HL}, \mathbf{X}_{HH}]
\end{equation}
where $\mathbf{X}_{LL}$ captures the coarse structural approximation, and the set $\{\mathbf{X}_{LH}, \mathbf{X}_{HL}, \mathbf{X}_{HH}\}$ encapsulates vertical, horizontal, and diagonal high-frequency details, respectively. 

By feeding this decomposed representation into the initial feature extraction layer, the network's input channels are pre-segregated by frequency. 
This imposes a strong Inductive Bias, the network no longer needs to expend additional capacity learning to identify high-frequency noise; it is explicitly provided. 
The subsequent encoder blocks can thus specialize-dedicating specific convolutional filters to preserve the structural integrity in the $\mathbf{X}_{LL}$ stream while aggressively suppressing noise in the high-frequency streams.

The core efficacy of the Wavelet-based approach lies in how the channel attention module interacts with the transformed data \cite{hou2021coordinate}. In our implementation, the attention mechanism is defined as:
\begin{equation}
\mathbf{y} = \mathbf{x} \cdot \sigma\left(\mathbf{W}_2 \cdot \delta(\mathbf{W}_1 \cdot \text{GAP}(\mathbf{x}))\right)
\end{equation}
where $\mathbf{x}$ represents the feature map, GAP denotes Global Average Pooling, $\mathbf{W}_1$ and $\mathbf{W}_2$ are learnable weight matrices, $\delta(\cdot)$ is the ReLU activation, and $\sigma(\cdot)$ is the sigmoid function.

In a standard spatial network, GAP aggregates spatial information, losing frequency specificity. 
However, in the wavelet domain, the channels carry explicit frequency meanings. 
Consequently, the channel attention mechanism evolves into a spectral gating mechanism. 
The global average of a high-frequency channel (e.g., derived from $\mathbf{X}_{HH}$) effectively measures the “noise energy” or “texture density” of the input. 

This allows the network to dynamically recalibrate features.
If the global statistics indicate a noise-dominated input (high energy in HH bands without corresponding structure in LL), the attention module drives the weights of these high-frequency channels toward zero ($\sigma(\cdot) \rightarrow 0$), effectively “gating” the speckle noise at the feature level.
Conversely, if strong activations are detected in $\mathbf{X}_{LL}$ and $\mathbf{X}_{HL}$ simultaneously (indicating a vertical edge), the attention mechanism boosts these channels to preserve the boundary.

This frequency-aware recalibration is mathematically unattainable in pure spatial networks without significantly deeper architectures.

The DWT operation downsamples the spatial resolution by a factor of 2, transforming the domain to $\frac{H}{2} \times \frac{W}{2}$. This geometric transformation has a profound impact on the Global Context Extractor (GCE) \cite{Ghasemabadi_2024}.

This expansion allows the CNN to capture long-range semantic dependencies and non-local self-similarities with significantly lower computational cost (GMACs). 
Since speckle noise is spatially local (granular) while biological structures often exhibit global continuity, the enlarged receptive field enables the GCE to differentiate between the two more robustly. 
The SCNet can leverage a wider context to verify if a high-frequency variation is an isolated speckle artifact or part of a continuous structural edge.

Classical wavelet denoising relies on “soft thresholding” or “shrinkage” functions to eliminate small coefficients (assumed to be noise) while keeping large coefficients (signal) \cite{Chen_2023_05}. The gating (SimpleGate) module in the code performs the operation:
\begin{equation}
f(\mathbf{x}) = \mathbf{x}_1 \odot \mathbf{x}_2
\end{equation}
where $\mathbf{x}$ is split into two halves. In the context of neural networks, this acts as a dynamic, data-driven activation function.

When applied to wavelet-domain features, the gating (SimpleGate) module functions as a learnable soft thresholding operator. The network learns to utilize one half of the channels ($\mathbf{x}_2$) as a gate for the other half ($\mathbf{x}_1$). 
Because the wavelet representation of natural images is sparse (most energy is compacted in few coefficients), the network can easily learn to set the gate values to zero for coefficients that correspond to speckle noise, while passing the coefficients that correspond to true signal. 
This mimics the mathematical properties of ideal wavelet denoising but with thresholds that are adaptively learned for the specific statistics of the speckle patterns \cite{chang2000adaptive, kopsinis2009development, bayer2019iterative}.

The enhanced performance of SCNet is attributed to the alignment between the mathematical properties of the Haar transform and the architectural components of the Gaze block. 
The DWT provides a sparse, frequency-segregated representation that allows the channel attention to act as a spectral filter and the gating to act as a non-linear denoiser. 
Furthermore, the inherent downsampling expands the effective receptive field of the GCE, enabling superior differentiation between noise and structure. 
Finally, the invertibility of the IDWT ensures that the reconstructed image maintains global consistency, free from the checkerboard artifacts often seen in purely strided-convolution decoding.

\subsection{Model distillation}\label{subsec17}

To enable compact neural networks to restore images of diverse material categories without encountering catastrophic interference, we developed the Multi-Teacher Knowledge Distillation (MT-KD) scheme. 
Five category-specific teacher models - all pre-trained on a single material database using the SCNet architecture (Extended Data Fig.\ref{fige4}a) - were frozen and deployed as a whole. 
During the distillation process, each training sample was routed to the corresponding teacher model based on its material metadata, generating category-specific soft targets (Extended Data Fig.\ref{fige4}b). 
This method avoids the performance degradation caused by direct multi-task optimization, while retaining the prior knowledge of specialized speckle removal.
The MT-KD strategy draws on the research ideas of lightweighting of super-resolution models by freezing the pre-trained teacher model to generate soft targets, the material-specific speckle suppression knowledge can be retained \cite{Zhang_2025_02,Chiang_2023_01}; the reduction of the computational cost of the student model follows the “performance - efficiency” balance principle in multi-task optimization, which has been proven to be able to reduce computational overhead while avoiding catastrophic forgetting \cite{Liu_2025_04}.

Student model architecture and feature adaptation. 
The student network is a simplified variant of SCNet, with a computational cost only 55\% of that of the teacher model. 
It is specifically designed for cross-category generalization. The distillation objective function combines three complementary terms:

\begin{equation}
\mathcal{L}_{\text{total}} = \alpha(\tau) \cdot \mathcal{L}_{\text{hard}} + \left(1 - \alpha(\tau)\right) \cdot \mathcal{L}_{\text{soft}}
\end{equation}
\begin{equation}
\mathcal{L}_{\text{hard}} = \text{Charbonnier}(I_{\text{output}}, I_{\text{GT}}) + \lambda_{\text{ssim}} \cdot (1 - \text{SSIM}(I_{\text{output}}, I_{\text{GT}}))
\end{equation}
\begin{equation}
\mathcal{L}_{\text{soft}} = \text{MSE}\left(\frac{I_{\text{output}}}{\tau}, \frac{I_{\text{teacher}}}{\tau}\right) = \frac{1}{n} \sum_{i=1}^{n} \left( \frac{I_{\text{output},i}}{\tau} - \frac{I_{\text{teacher},i}}{\tau} \right)^2
\end{equation}

Among them, $\mathcal{L}_{\text{hard}}$ combines the Charbonnier function with the structural similarity loss function; $\mathcal{L}_{\text{soft}}$ is designed for the regression task, which aims to minimize the mean square error of the temperature-scaled output; and $\alpha(\tau)$ denotes the temperature scaling coefficient.
The dynamic weight coefficient $\alpha(\tau)$ decreases linearly from 0.70 to 0.30 within the training period (500,000 iterations), gradually shifting the focus of supervision from the true labels to the teacher guidance. Among the equation (6) $\lambda_{\text{ssim}}$ is 0.2. Among the equation (7) $\tau$ is 10, Which is controlling the smoothness of the transmission of teacher signals

\begin{align}
\mathcal{L}_{\text{Charbonnier}}(y, \bar{y}) &= \frac{1}{N} \sum_{i=1}^{N} \rho(y - \bar{y}), \\
\rho(x) &= \sqrt{x^2 + \varepsilon^2},
\end{align}

The above equation (8) represents the Charbonnier loss function, where $y$ is the predicted value, $\bar{y}$ is the ground truth value, $N$ is the total number of pixels, and $\varepsilon$ is a small constant (typically set to $1 \times 10^{-6}$) to ensure numerical stability.

The training process involved 500,000 iterations. 
The total batch size on a single NVIDIA RTX 4090 GPU was 4. 
The teacher model remained frozen. Gradient updates used the AdamW optimizer with parameters $\beta_1 = 0.9$, $\beta_2 = 0.999$.
The learning rate followed a cosine annealing restart schedule: two consecutive cycles of 250,000 iterations each, with initial peak values of $8 \times 10^{-5}$ and $1 \times 10^{-5}$, decreasing to the lower limit of $3 \times 10^{-7}$ and $1 \times 10^{-8}$, and each cycle restarting the weights at $1.0$. 
Gradient clipping (with an $L_2$ norm upper limit of $1.0$) stabilized the early training.

\subsection{Training details}\label{subsec18}
Individual model training was executed on a single NVIDIA GeForce RTX 4090 GPU to demonstrate computational accessibility. 
For the comprehensive evaluation including benchmarking and ablation studies, we utilized a heterogeneous compute environment consisting of four NVIDIA GeForce RTX 4090s, one NVIDIA RTX 5880 Ada, one NVIDIA RTX A5000 and one NVIDIA RTX A4500.

The SCNet model was implemented using the PyTorch framework. 
The training process was carried out on an NVIDIA GeForce RTX 4090 GPU or an NVIDIA RTX 5880 Ada. 
The total number of training iterations was set to 400,000. Optimize for 200,000 steps using frequency domain and spatial domain loss functions respectively.
The batch size was set to 4. The Adam optimizer was used for gradient updates, with parameters $\beta_1 = 0.9$, $\beta_2 = 0.999$. 
The learning rate followed a cosine annealing schedule, starting from an initial value of $1 \times 10^{-4}$ and gradually decreasing to a minimum value of $1 \times 10^{-6}$. 

During the training process, to maintain the physical characteristics of the original data, especially the fine material textures, no data augmentation techniques were applied during the training process. 
All input images were cropped into single-channel image blocks of 220$\times$220 pixels, and their pixel values were normalized to the [0, 1] range before being input into the network.

The quantitative accuracy of the image restoration results was evaluated using established image quality metrics. 
Details are provided in Supplementary Information S4.
The PSNR, RMSE and SSIM are proposed to measure the pixel-level and structure-level similarities between a restored image $\tilde{y}$ and a GT image $y$ \cite{ma2024pretraining, wang2004image}.

\subsection{Design of loss function}\label{subsec19}

The core of our training strategy is a two-stage, coarse-to-fine optimization paradigm, aiming to systematically address the fundamental contradiction inherent in speckle removal, namely the strong noise suppression and the preservation of fine details. 
The training process and loss function are shown in Fig.\ref{fig1}c .

In the initial stage, the inherent characteristics of wavelet transformation are utilized to decouple the problem. 
The input speckle image is first decomposed into four frequency sub-bands “LL, LH, HL and HH”, using an unlearnable two-dimensional discrete wavelet transform (DWT, Haar basis). 
This decomposition effectively separates the high-frequency particle components of the speckle noise (mainly in the LH, HL and HH detail sub-bands) from the low-frequency structural information of the image's base layer (concentrated in the LL approximation sub-band).
(Fig.\ref{fig1}c) As shown in the following formula.
\begin{equation}
\mathcal{L}_{\text{total}} = \sum_{b \in \{LL, LH, HL, HH\}} w_b \cdot \mathcal{L}_{\text{Charbonnier}}(\text{DWT}(P)_b, \text{DWT}(T)_b)
\end{equation}

The total loss $\mathcal{L}_{\text{total}}$, is calculated as the weighted sum of individual loss terms applied across the four frequency sub-bands of the predicted tensor $P$ and the target tensor $T$. Both $P$ and $T$ are first decomposed using a single-level 2D Haar Discrete Wavelet Transform, denoted as $\text{DWT}(\cdot)$. This operation yields four components indexed by $b \in \{LL, LH, HL, HH\}$, which represent the approximation ($LL$) and the horizontal, vertical, and diagonal details ($LH, HL, HH$), respectively. For each sub-band $b$, a specific loss function $\mathcal{L}_{\text{Charbonnier}}$ is applied to its corresponding components from the prediction, $\text{DWT}(P)_b$, and the target, $\text{DWT}(T)_b$. 
Finally, each sub-band's loss is multiplied by a scalar coefficient, $w_b$, which weights its relative importance, and these weighted losses are summed to produce $\mathcal{L}_{\text{total}}$.
The Charbonnier loss (a smoothed approximation of the L1 norm) was chosen because it is robust to the high-intensity outliers that are commonly present in the detail subbands caused by speckle “hotspots”. 
This can prevent the network from excessively suppressing the true edge information - a common defect when using L2-based losses - thereby enabling powerful and targeted denoising in the frequency domain.

After the speckle removal in the first stage, the optimization objective shifts to high-fidelity fine-tuning in the spatial domain \cite{Xie_2023_10}. 
During this stage, we fine-tune the entire end-to-end model, from the initial DWT layer to the IDWT layer that reconstructs the final spatial image. 
We employed a hybrid loss function that synergistically combines a perceptual quality term and a pixel fidelity term:

\begin{align}
\mathcal{L}_{\text{Hybrid}} &= \alpha(\tau) \cdot \left( \beta - \mathcal{L}_{\text{PSNR}} \right) + \alpha(\tau) \cdot \mathcal{L}_{\text{SCIM}}, \\
\alpha(\tau) &= \left( \frac{\mathcal{L}_{\text{PSNR}}}{\mathcal{L}_{\text{SCIM}} + \varepsilon} \right)
\end{align}

Here, the SCIM (Stochastic Consistency Invariance Measure) loss function directly encourages structural consistency with the true values, which is crucial for preserving the morphology of biological structures in endoscopic imaging. Meanwhile, the PSNR loss function enforces strict pixel-level accuracy, serving as a regularization term to prevent the model from “imagining” reasonable but incorrect details.
In this equation(11-12), $\beta$ is set to 50, $\varepsilon$ is a small constant (typically set to ${10^{-6}}$) to ensure numerical stability, and $\tau$ is the training iteration number. $\alpha(\tau)$ is a dynamic weight coefficient. Through the ratio of $\mathcal{L}_{\text{PSNR}}$ to $\mathcal{L}_{\text{SCIM}}$ (supplemented by a small perturbation $\varepsilon$ to avoid singularities), it adaptively establishes a balance mechanism between the pixel fidelity loss and structural similarity loss. 
This enables refined weight allocation for the dual supervision objectives in the hybrid loss $\mathcal{L}_{\text{Hybrid}}$, driving the model to intelligently trade off the optimization directions of pixel-level accuracy and structural-level consistency during training.

This phased approach logically guides the learning trajectory of the network: Firstly, by learning the physical priors of speckle noise in the frequency domain, a clean and well-structured baseline is established; then, in the spatial domain, the perception and pixel-level accuracy are finely optimized.

\section{Data availability}\label{sec6}
Part of the test datasets used for model evaluation in this study, have been made publicly available at https://doi.org/10.6084/m9.figshare.30995104.
Due to the substantial volume of the complete training dataset (comprising synchronized image pairs of plastic, paper, metal, plant leaves, and rabbit heart and kidney tissues), it is not currently feasible to host the full dataset in a public repository.
The complete datasets are available from the corresponding author upon reasonable request.

\section{Code availability}\label{sec7}
All the code of the models in our algorithm can be accessed online through the GitHub repository. 
This repository provides the training and testing codes for SCNet, SCNet-MoE and distillation.
The code of SCNet, SCNet-MoE and distillation are publicly available at https://github.com/zalbert-op/SCNet.
The code of ECCNet is publicly available at https://github.com/zalbert-op/ECCNet.


\bibliography{p1}

\section{Acknowledgments}\label{sec8}
X.R.Z. and Y.D. acknowledge support from the Hangzhou Joint Fund of the Zhejiang Provincial Natural Science Foundation of China (Grant No. LHZY24F030002) and funding from the Hangzhou Institute for Advanced Study, UCAS. Y.R.Z. and Y.Y. acknowledge support from the National Science Foundation for Distinguished Young Scholars of China (52325106), National Key R\&D Program of China (2020YFA0211100), the National Science Foundation of China (51922077), Program of Shanghai Academic Research Leader(23XD1402900), and the Foundation of National Facility for Translational Medicine (Shanghai) (TMSK-2020-012). 
T.\v{C}. acknowledges support from the European Research Council (724530), the Ministry of Education, Youth and Sports of the Czech Republic (CZ.02.1.01/0.0/0.0/15 003/0000476), the European Regional Development Fund (LM2018129), the European Union’s H2020-RIA (101016787). 
Y.D. further acknowledges Dr. Ivo T. Leite, Dr. Dirk E. Boonzajer Flaes, Dr. Andr\'e Gomes and Dr. Sergey Turtaev for fruitful discussion at early stage of this work.

\section{Author contributions}\label{sec9}
T.\v{C}. and Y.D. initiated and conceived the background for the presented technology. 
X.R.Z. conceptualized and designed the deep learning series models. 
X.R.Z. and Y.D. built the experimental setup with support from T.\v{C}.. 
X.R.Z. performed all the data acquisition and training. 
Y.R.Z. carried out the collection of relevant animal organs and performed surgeries. 
X.R.Z. and P.F.L. prepared figures. 
X.R.Z., F.Y., and Y.D. analyzed the results. 
Y.Y. oversaw animal ethics compliance. 
Y.Y., T.\v{C}. and Y.D. secured the funding. 
X.R.Z. drafted the initial manuscript. 
Y.D. supervised the research and finalized the manuscript with contributions from all authors.

\section{Competing interests}\label{sec10}
T.\v{C}. is co-founder of company DeepEn GmBH. The other authors declare no competing interests.

\newpage
\setcounter{figure}{0} 
\renewcommand{\figurename}{Extended Data Fig.}
\renewcommand{\theHfigure}{ED\arabic{figure}}

\begin{figure}[htbp!]
\centering
\includegraphics[width=1\textwidth]{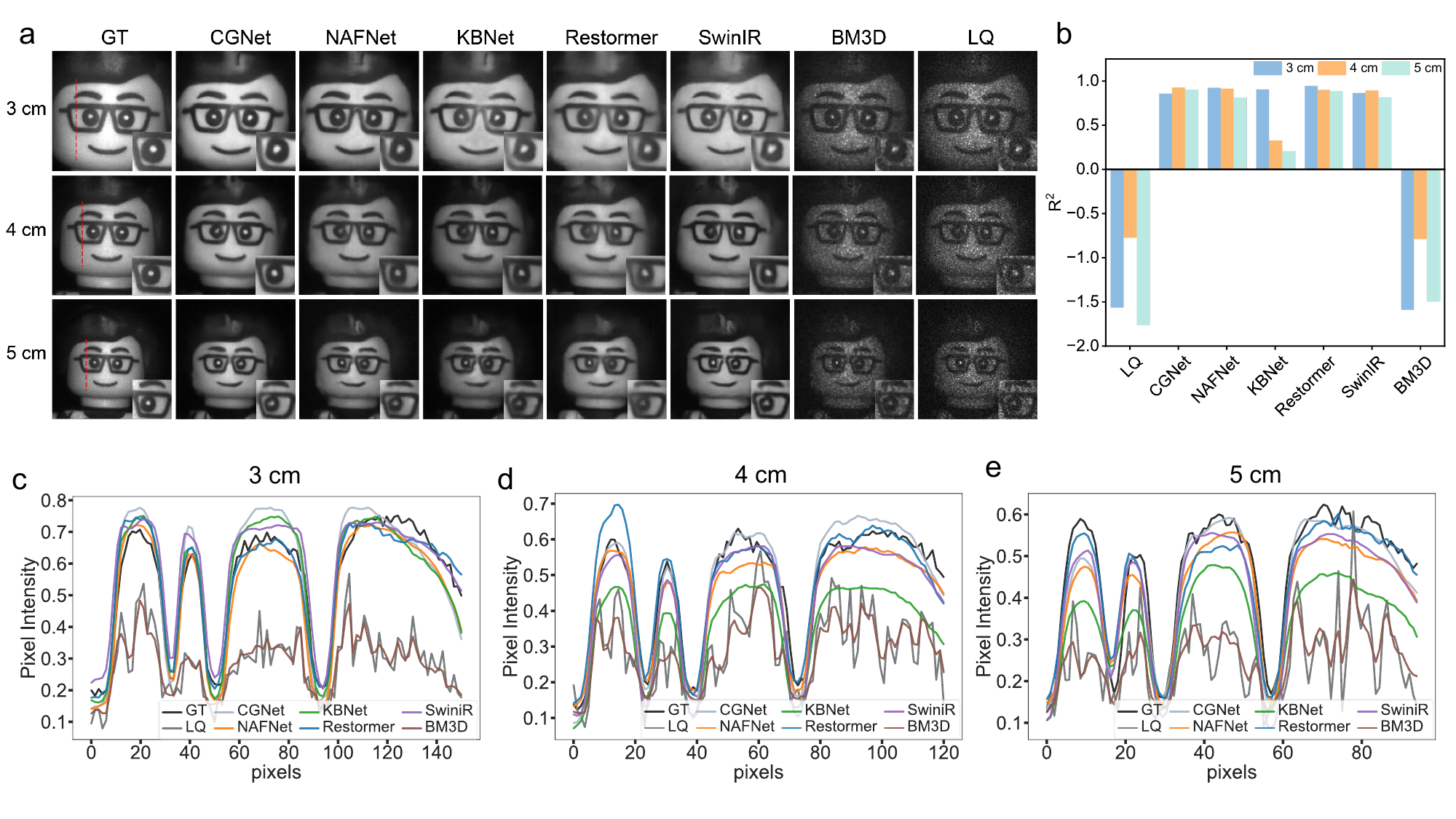}
\refstepcounter{figure}
\caption*{\textbf{Extended Data Fig.1}\textbar\textbf{Quantitative evaluation of speckle suppression performance of plastic samples under different models and different imaging distances.} 
\textbf{a,} Comparison of the speckle removal effect of the Lego minifigure's facial image by the GT and different models at working distances of 3 cm, 4 cm, and 5 cm.
\textbf{b,} Comparison of the goodness of fit (R²) of each method's linear intensity profile with GT at different distances.
\textbf{c-e,} Comparison of the linear intensity distribution curves along the selected line of each model at distances of 3 cm, 4 cm, and 5 cm with GT.}
\addcontentsline{lof}{figure}{}
\label{fige1}
\end{figure}

\begin{figure}[htbp!]
\centering
\includegraphics[width=1\textwidth]{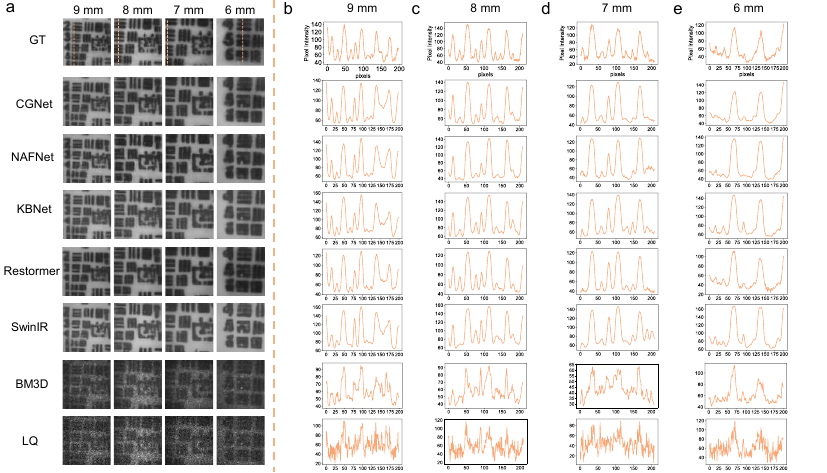}
\refstepcounter{figure}
\caption*{\textbf{Extended Data Fig.2}\textbar\textbf{Evaluation of the recovery performance of paper-based USAF resolution targets under different model methods and different imaging distances.} 
\textbf{a,} Comparison of imaging performance for the USAF resolution target at four working distances (9 mm, 8 mm, 7 mm, and 6 mm), including GT, outputs from each comparison model, and the LQ (imaging field of view correspond to those in Fig.\ref{fig2}b-e in the main text).  
\textbf{b-e,} Linear intensity distribution curves of GT, model results, and LQ along selected lines at distances of 9 mm (b), 8 mm (c), 7 mm (d), and 6 mm (e).}
\addcontentsline{lof}{figure}{}
\label{fige2}
\end{figure}

\begin{figure}[htbp!]
\centering
\includegraphics[width=1\textwidth]{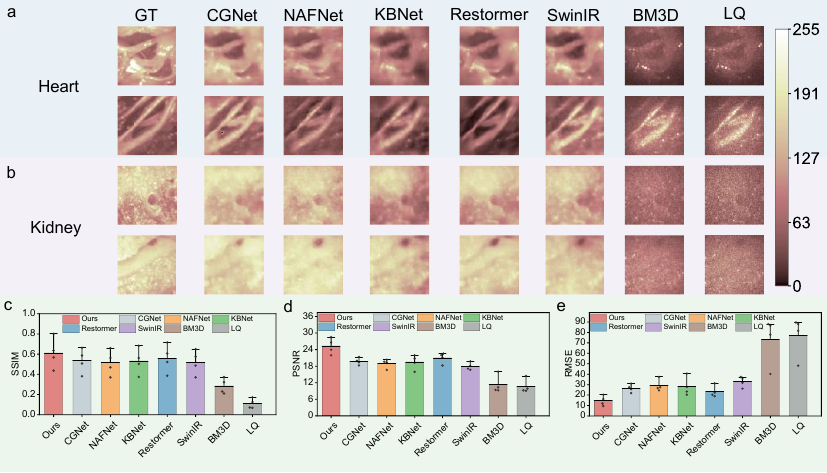}
\refstepcounter{figure}
\caption*{\textbf{Extended Data Fig.3}\textbar\textbf{Comparison of speckle suppression performance across different methods on biological tissues.} 
\textbf{a,} GT, LQ, and restoration results from other comparison models on a rabbit heart sample (same sample as Fig.\ref{fig5}a in the main text).  
\textbf{b,} GT, LQ, and restoration results from other comparison models for rabbit kidney sample (same samplea in Fig.\ref{fig5}b in the main text). 
\textbf{c-e,} Quantitative comparison of SSIM, PSNR, and RMSE metrics for each method on the above restoration results; error bars indicate outlier ranges.}
\addcontentsline{lof}{figure}{}
\label{fige3}
\end{figure}

\begin{figure}[htbp!]
\centering
\includegraphics[width=1\textwidth]{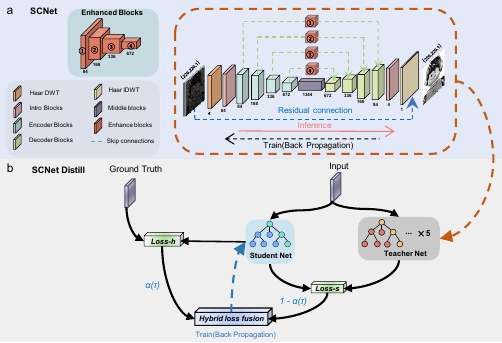}
\refstepcounter{figure}
\caption*{\textbf{Extended Data Fig.4}\textbar\textbf{Schematic diagram of the SCNet model architecture and distill model.}
\textbf{a,} Core network structure diagram of the SCNet deep learning model, showing its multi-scale encoding-decoding architecture and key module composition.
\textbf{b,} Schematic diagram of the knowledge distillation process of the SCNet model, demonstrating the information transmission and lightweight student network construction process under the multi-teacher knowledge distillation framework.}
\addcontentsline{lof}{figure}{}
\label{fige4}
\end{figure}

\begin{table}[htbp]
\centering
\renewcommand{\arraystretch}{2.5}
\setlength{\tabcolsep}{10pt}
\caption*{\textbf{Extended Data Table 1 \textbar\ Quantitative comparison of mean image restoration performance across five material test datasets}}
\begin{tabular}{l|c|c|c}
\hline
\textbf{Model} & \textbf{SSIM $\uparrow$} & \textbf{PSNR $\uparrow$} & \textbf{RMSE $\downarrow$} \\
\hline
SCNet (MSE-Loss) & 0.721 & 26.720 & 13.046 \\
\hline
SCNet (CDPO-Loss)    & 0.739 & 28.208 & 11.230 \\
\hline
SCNet (CLAHE)   & 0.739 & 28.354 & 10.890 \\
\hline
SCNet (Distill) & 0.700 & 24.178 & 17.754 \\
\hline
CGNet           & 0.682 & 22.726 & 20.872 \\
\hline
NAFNet          & 0.675 & 22.698 & 21.016 \\
\hline
KBNet           & 0.681 & 22.616 & 21.458 \\
\hline
Restormer       & 0.690 & 23.948 & 21.030 \\
\hline
SwinIR          & 0.688 & 23.360 & 19.436 \\
\hline
BM3D            & 0.408 & 13.884 & 55.396 \\
\hline
LQ              & 0.190 & 13.234 & 59.088 \\
\hline
\end{tabular}

\medskip 
{\footnotesize The specific formulas for SSIM, PSNR, and RMSE are provided in the supplementary materials. 
The arrows adjacent to these three values indicate whether higher values are better or lower values are better.
SCNet (MSE-Loss) indicates the model that does not use our curriculum learning loss function, and only adopts MSE loss. SCNet (CDPO-Loss) refers to the model using the loss function of our curriculum learning method. 
SCNet (CLAHE) is built upon SCNet (CDPO-Loss) with the addition of the CLAHE histogram equalization algorithm; details of this algorithm are provided in the supplementary materials. 
SCNet (Distill) denotes the distilled model derived from SCNet (CDPO-Loss) through knowledge distillation.
}
\label{tab:quantitative_comparison}
\end{table}

\begin{table}[htbp]
\centering
\renewcommand{\arraystretch}{2.5}
\setlength{\tabcolsep}{12pt}
\caption*{\textbf{Extended Data Table 2 \textbar\ Quantitative comparison of computational efficiency}}
\begin{tabular}{l|c|c}
\hline
\textbf{Model} & \textbf{FPS $\uparrow$} & \textbf{GMACs $\downarrow$} \\
\hline
SCNet (MSE-Loss) & 35.6 & 63.43 \\
\hline
SCNet (CDPO-Loss)    & 35.6 & 63.43 \\
\hline
SCNet (CLAHE)   & 35.6 & 63.43 \\
\hline
SCNet (Distill) & 60.0 & 34.67 \\
\hline
CGNet           & 35.4 & 183.29 \\
\hline
NAFNet          & 35.2 & 183.29 \\
\hline
KBNet           & 7.8  & 228.72 \\
\hline
Restormer       & 19.0 & 151.73 \\
\hline
SwinIR          & 8.0  & 257.29 \\
\hline
\end{tabular}

\medskip 
{\footnotesize FPS and GMACs are used to evaluate the computational efficiency. 
Higher FPS and lower GMACs indicate better efficiency.
The specific calculation methods for FPS and GMACs are provided in the supplementary materials.}
\label{tab:computational_efficiency}
\end{table}

\end{document}